\documentclass[aps,prl,twocolumn,superscriptaddress]{revtex4}

\usepackage{graphicx,amsmath,bm,textcomp}
\usepackage{color}
\usepackage{textgreek}
\usepackage{amssymb,amsfonts}
\newcommand{\unit}[1]{\ensuremath{\, \mathrm{#1}}}

\begin{document}

\title{Cadherin-Based Intercellular Adhesions Organize \\ Epithelial Cell-Matrix Traction Forces}

\author{Aaron~F.~Mertz}
\affiliation{Department of Physics, Yale University, New Haven, Connecticut 06520, USA}
\author{Yonglu~Che}
\affiliation{Department of Physics, Yale University, New Haven, Connecticut 06520, USA}
\affiliation{Department of Molecular, Cellular, and Developmental Biology, Yale University, New Haven, Connecticut 06520, USA}
\author{Shiladitya~Banerjee}
\affiliation{Department of Physics, Syracuse University, Syracuse, New York 13244, USA}
\author{Jill~Goldstein}
\affiliation{Department of Molecular, Cellular, and Developmental Biology, Yale University, New Haven, Connecticut 06520, USA}
\author{Kathryn~R.~Rosowski}
\affiliation{Department of Molecular, Cellular, and Developmental Biology, Yale University, New Haven, Connecticut 06520, USA}
\author{Carien~M.~Niessen}
\affiliation{Department of Dermatology, Center for Molecular Medicine, Cologne Excellence Cluster on Cellular Stress Responses in Aging-Associated Diseases, University of Cologne, 50931 Cologne, Germany}
\author{M.~Cristina~Marchetti}
\affiliation{Department of Physics, Syracuse University, Syracuse, New York 13244, USA}
\affiliation{Syracuse Biomaterials Institute, Syracuse University, Syracuse, New York 13244, USA}
\author{Eric~R.~Dufresne}
\email[]{eric.dufresne@yale.edu}
\affiliation{Departments of Mechanical Engineering \& Materials Science, Chemical \& Environmental Engineering, and Cell Biology, Yale University, New Haven, Connecticut 06520, USA}
\affiliation{Department of Physics, Yale University, New Haven, Connecticut 06520, USA}
\author{Valerie~Horsley}
\email[]{valerie.horsley@yale.edu}
\affiliation{Department of Molecular, Cellular, and Developmental Biology, Yale University, New Haven, Connecticut 06520, USA}

\date{\today}

\begin{abstract}

Cell--cell and cell-matrix adhesions play essential roles in the function of tissues.  
There is growing evidence for the importance of crosstalk between these two adhesion types, yet little is known about the impact of these interactions on the mechanical coupling of cells to the extracellular-matrix (ECM).  
Here, we combine experiment and theory to reveal how intercellular adhesions modulate forces transmitted to the ECM. 
In the absence of cadherin-based adhesions, primary mouse keratinocytes within a colony appear to act independently, with significant traction forces extending throughout the colony.
In contrast, with strong cadherin-based adhesions, keratinocytes in a cohesive colony localize traction forces to the colony periphery. 
Through genetic or antibody-mediated loss of cadherin expression or function, we show that cadherin-based adhesions are essential for this mechanical cooperativity.
A minimal physical model in which cell--cell adhesions modulate the physical cohesion between contractile cells is sufficient to recreate the spatial rearrangement of traction forces observed experimentally with varying strength of cadherin-based adhesions.
This work defines the importance of cadherin-based cell--cell adhesions in coordinating mechanical activity of epithelial cells and has implications for the mechanical regulation of epithelial tissues during development, homeostasis, and disease.
\bigskip
\end{abstract}

\maketitle

Mechanical interactions of individual cells have a crucial role in the spatial organization of tissues~\cite{Zallen-2007,Farhadifar-2007} and in embryonic development~\cite{Gorfinkiel-2009,Mammoto-2010,Goehring-2011}.
The mechanical cooperation of cells is evident in dynamic processes such as flow-induced alignment of vascular endothelial cells~\cite{Davies-1995} and muscle contraction~\cite{Iwamoto-2000}.
However, mechanical interactions of cells within a tissue also affect the tissue's static mechanical properties including elastic modulus~\cite{Engler-2004}, surface tension~\cite{Foty-1994}, and fracture toughness~\cite{Gokgol-2012}.
Little is known about how these tissue-scale mechanical phenomena emerge from interactions at the molecular and cellular levels~\cite{Humphrey-2008}.

Tissue-scale mechanical phenomena are particularly important in developmental morphogenesis~\cite{Martin-2010}, homeostasis~\cite{Vaezi-2002}, and wound healing~\cite{Fenteany-2000} in epithelial tissues.
Cells exert mechanical force on each other at sites of intercellular adhesion, typically through cadherins~\cite{Liu-2010,Borghi-2012}, as well as on the underlying extracellular matrix (ECM) through integrins~\cite{Balaban-2001,Wang-2007,Sabass-2008}.
Cadherin-based adhesions can alter physical aspects of cells such as the surface tension of cellular aggregates~\cite{Foty-2005} and the spreading~\cite{Ladoux-2010} and migration~\cite{Borghi-2010} of single cells adherent to cadherin-patterned substrates.
Integrity of intercellular adhesions may also contribute to metastatic potential~\cite{Onder-2008}.
We and others have shown that epithelial cell clusters with strong cell--cell adhesions exhibit coordinated mechanical behavior over length scales much larger than a single cell~\cite{Trepat-2009,Maruthamuthu-2011,Mertz-2012}.
Several studies have implicated crosstalk between cell-ECM and cell--cell adhesions~\cite{Tsai-2009,McCain-2012} that can be modulated by actomyosin contractility~\cite{deRooij-2005}.
Recent data suggest that integrin-mediated adhesions can modulate the composition~\cite{Yamada-2007,Tseng-2012} and tension~\cite{MartinezRico-2010,Maruthamuthu-2011,Weber-2011} of cell--cell junctions.
While cadherins have been shown to modify local traction forces~\cite{Jasaitis-2012} and monolayer contractility~\cite{Krishnan-2011}, the effects of intercellular adhesions on the spatial organization of cell-ECM forces remain unexplored.

In this paper, we address the impact of intercellular adhesions on cell-ECM traction forces in colonies of primary mouse keratinocytes.
We measure tractions of colonies of keratinocytes before, during, and after formation of cadherin-mediated intercellular adhesions.
As cadherin-dependent junctions form, there is dramatic rearrangement of cell-ECM traction forces from a disorganized, punctate distribution underneath the colony to an organized concentration of force at the colony periphery.
Through perturbations of cadherin-based adhesions, we demonstrate an essential role for cadherin in organizing cell-matrix mechanics.
Finally, the spatial reorganization of cell-matrix forces is reproduced by a minimal physical model of a cell colony as two-dimensional objects connected by springs and adherent to a soft substrate.
While downstream signaling pathways regulate responses to cadherin-based-junction formation, our experimental data and physical model suggest that the simple physical cohesion created by intercellular adhesions is sufficient to organize traction forces.
These results have implications for intercellular adhesions' role in the mechanical relationship of tissues to their surroundings and the emergence of tissues' bulk material properties.

\section{Results}

\subsection{Traction Stresses Dynamically Reorganize in High-Calcium Medium}
To investigate the relationship between cadherin-based intercellular adhesions and cell-matrix traction stresses, we induced the formation of cadherin-based adhesions in primary  mouse keratinocytes by elevating extracellular-calcium concentrations.
In low-calcium medium, keratinocytes plated at low density proliferated into colonies of cells with weak cell--cell interactions.
Exposing keratinocytes to high-calcium medium resulted in formation of cadherin-based cell--cell adhesions after $6$--$12 \unit{h}$.

\begin{figure}[hbt]
\includegraphics[width=0.43\textwidth]{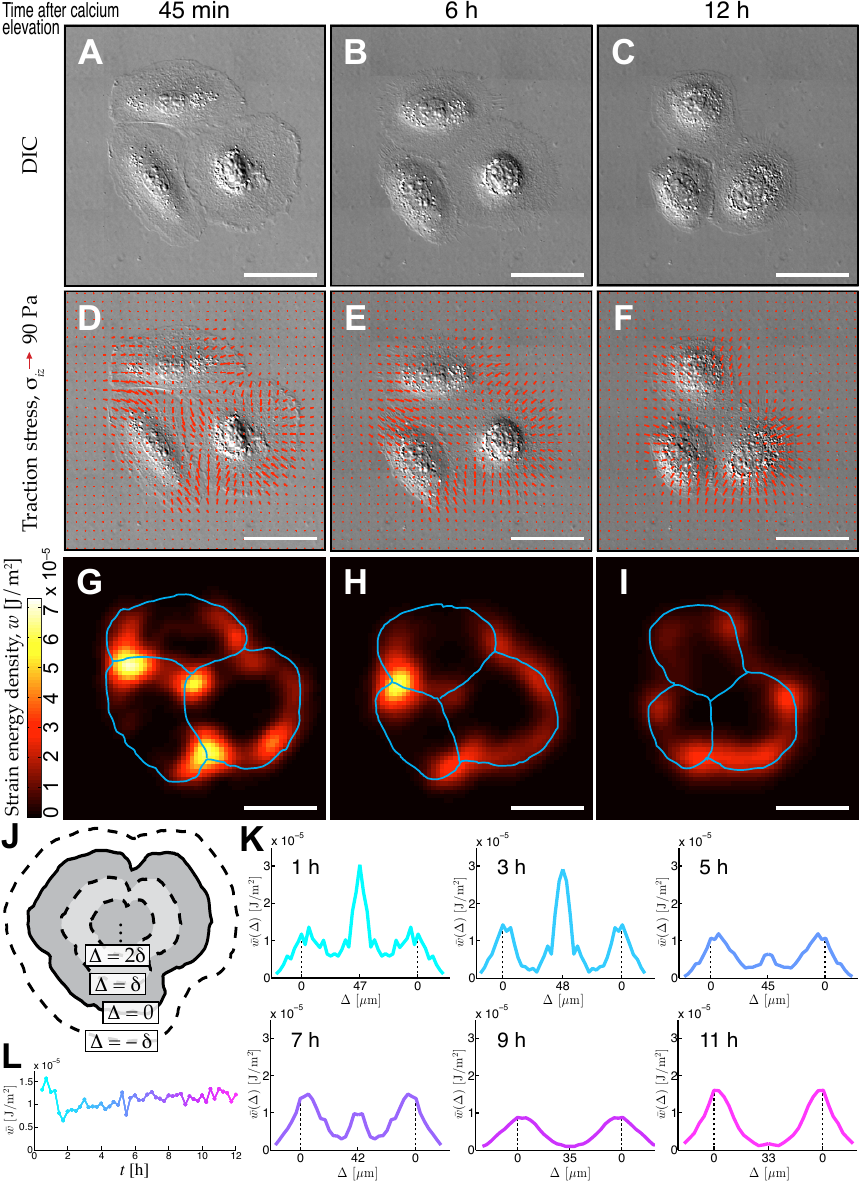}
\caption{\label{fig:Fig1} \textit{Traction stresses dynamically reorganize in high-calcium medium.}
(\textit{A}--\textit{C}) DIC images of a three-cell colony at $45 \unit{min}$ (\textit{A}), $6 \unit{h}$ (\textit{B}), and $12 \unit{h}$ (\textit{C}) after calcium elevation.
(\textit{D}--\textit{F}) Distribution of in-plane traction stresses (red arrows) for cell colony at timepoints in \textit{A}--\textit{C} overlaid on DIC images.
For clarity, one-quarter of calculated traction stresses are shown.
(\textit{G}--\textit{I}) Distribution of strain energy density, $w$, for cell colony at timepoints in \textit{A}--\textit{C}.
Blue lines mark individual cell boundaries.
(\textit{J}) Schematic for calculating azimuthal-like averages for strain energy.
Colony outline is eroded inward by distance, $\Delta$, in discrete steps, $\delta$, until entire colony area has been covered.
Average strain energy density is then calculated for each concentric, annular-like region.
(\textit{K}) Strain energy profiles for three-cell colony at six timepoints after calcium elevation.
Solid colored lines represent colony's average strain energy density as a function of distance, $\Delta$, from colony edge.
Each profile is mirrored about $\Delta \approx R$, the effective colony radius.
Colony periphery ($\Delta = 0$) is indicated by dashed vertical black lines.
Strain energy at $\Delta <0$ corresponds to regions outside colony periphery.
(\textit{L}) Average strain energy density for entire colony at $15 \unit{min}$ intervals from $30 \unit{min}$ to $12 \unit{h}$ after calcium elevation.
Plot colors in \textit{K} and \textit{L} are scaled according to time, $t$, after calcium elevation, from cyan at $t=0$ to magenta at $t = 12 \unit{h}$.
Scale bars in \textit{A}--\textit{I} represent $50 \unit{\mu m}$.
}
\end{figure}

We quantified the effect of cell--cell adhesions on cell-matrix forces using traction force microscopy (TFM)~\cite{Dembo-1999}.
We plated keratinocytes onto fibronectin-coated, elastic silicone gel coupled to glass.
To quantify gel deformation due to cell-ECM traction force, we imaged fluorescent beads embedded in the silicone gel and measured the beads' displacements relative to their positions after removing the cells with \hbox{proteinase K}.
We calculated in-plane traction stresses, $\sigma_{iz}$, from bead displacements and the substrate's elastic properties~\cite{delAlamo-2007,Xu-2010} (\textit{Materials \& Methods}).
 
Over $12 \unit{h}$ in high-calcium medium, keratinocytes developed cell--cell junctions~\cite{Okeefe-1987} and contracted~\cite{Thornton-2008} (Fig.\ \ref{fig:Fig1} \textit{A--C}).
Prior to adhesion formation, in-plane traction stresses emanated from both the colony periphery and the interior junction of the three cells in a colony.
Forces at the colony periphery pointed radially inward, while interior forces pointed in various directions (Fig.\ \ref{fig:Fig1}\textit{D}).
During the timecourse, traction stress in the middle of the colony gradually weakened (Fig.\ \ref{fig:Fig1}\textit{E}), and by $12 \unit{h}$ after calcium elevation, interior traction stress all but disappeared (Fig.\ \ref{fig:Fig1}\textit{F}).

From substrate displacement and traction stresses, we calculated the strain energy density, $w$, the mechanical work per unit area performed by the colony to deform the substrate~\cite{Butler-2001} (\textit{Materials \& Methods}).
Shortly after calcium elevation, high strain energy was localized both underneath and at the periphery of the colony (Fig.\ \ref{fig:Fig1}\textit{G}).
$12 \unit{h}$ after calcium elevation, strain energy was limited to the colony edge (Fig.\ \ref{fig:Fig1}\textit{I}).

To quantify these spatial changes, we calculated azimuthal-like averages of strain energy during the timecourse.
We eroded the colony outline inward by distance, $\Delta$, in discrete steps, $\delta$, until the entire colony area was covered (Fig.\ \ref{fig:Fig1}\textit{J}).
We calculated the average strain energy, $\bar{w}(\Delta)$, in each of these concentric, annular-like regions and plotted it as a function of distance from the colony edge, $\Delta$ (Fig.\ \ref{fig:Fig1}\textit{K}).
During the first $3 \unit{h}$ after calcium elevation, three peaks exist in the strain-energy profiles, corresponding to localization of strong strain energy at the colony periphery ($\Delta=0$) and center.
Between $5$ and $9 \unit{h}$, the center strain-energy peak diminishes and disappears, and high strain energy is only at the colony periphery.
We measured some strain energy outside the colony ($\Delta<0$) due to the finite spatial resolution of our implementation of TFM.

Although strain-energy localization changed after calcium elevation, the colony's overall average strain energy density was relatively consistent during the timecourse (Fig.\ \ref{fig:Fig1}\textit{L}).
Hotspots of strong strain energy (yellow regions in Fig.\ \ref{fig:Fig1}\textit{G}) were no longer present by the end of the experiment (Fig.\ \ref{fig:Fig1}\textit{I}), but overall average strain energy density was compensated by a decrease in colony area.

\subsection{Traction Stresses Systematically Reorganize in High-Calcium Medium}
To probe how intercellular adhesions alter traction forces across a large range of colony geometrical size and cell number, we analyzed the magnitude and localization of traction force in 32 keratinocyte colonies in low-calcium medium and 29 keratinocyte colonies after $24 \unit{h}$ in high-calcium medium.
A total of 117 low-calcium cells and 150 high-calcium cells comprised these colonies, each containing 2--27 cells, and spanned a geometrical dynamic range of nearly a factor of 100 in spread area.

\begin{figure*}[hbt]
\includegraphics[width=0.87\textwidth, clip=true]{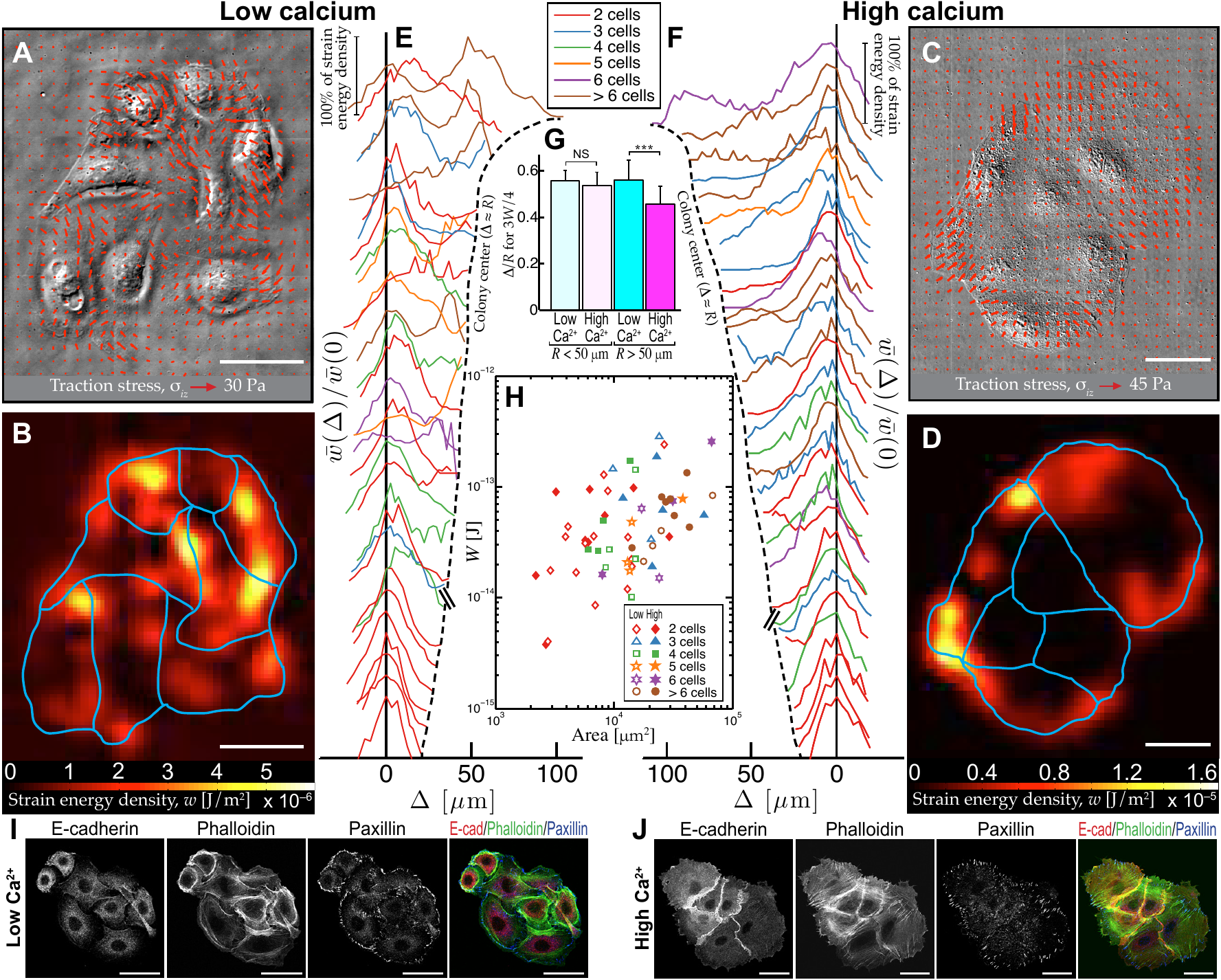}
\caption{\label{fig:Fig2} \textit{Traction stresses systematically reorganize in high-calcium medium.}
(\textit{A}) Distribution of in-plane traction stresses (red arrows) of an eight-cell wildtype colony in low-calcium medium overlaid on DIC image of colony.
For clarity, one-ninth of calculated traction stresses are shown.
(\textit{B}) Strain energy distribution, $w$, of low-calcium colony in \textit{C} with individual cell outlines in blue.
(\textit{C}) Distribution of traction stresses (red arrows) of a six-cell wildtype colony in high-calcium medium for $24 \unit{h}$ overlaid on DIC image of colony.
For clarity, one-ninth of calculated traction stresses are shown.
(\textit{D}) Strain energy distribution, $w$, of high-calcium colony in \textit{E}, with individual cell outlines in blue.
(\textit{E}) Strain energy profiles for $n=32$ low-calcium colonies.
(\textit{F}) Strain energy profiles for $n=29$ high-calcium colonies.
In \textit{E} and \textit{F}, each solid curve represents a colony's average strain energy density as a function of distance, $\Delta$, from colony edge.
Each profile terminates where inward erosion covers entire colony area, at $\Delta \approx R$, the effective colony radius, indicated by dashed line.
The erosion is defined in legend of Fig.\ \ref{fig:Fig1}\textit{J}.
Average strain energy is normalized to value at colony periphery, $\bar{w}(0)$, giving each colony the same height on the graphs, indicated by the vertical scale bar.
For clarity, profiles are spaced vertically according to colony size, with profiles for larger colonies (terminating at larger values of $\Delta$) appearing higher up the $y$ axis.
Profile colors correspond to colony cell number given in the legend.
(\textit{G}) Quantification of relative distance from colony periphery ($\Delta/R$) corresponding to 75\% of total strain energy, $3W/4$, in colonies in low- or high-calcium medium.
Small colonies ($R<50 \unit{\mu m}$, below hash marks in \textit{E} and \textit{F}), in low- ($n=8$) or high-calcium ($n=8$) medium showed no significant difference, whereas large ($R>50 \unit{\mu m}$) low-calcium colonies ($n=24$) had significantly more strain energy closer to colony center than large high-calcium colonies ($n=21$).
Statistical significance between low- and high-calcium populations is indicated by asterisks ($p<0.001$).
Error bars indicate one standard deviation.
(\textit{H}) Relationship between total strain energy, $W$, and area, $A$, of colonies in low- and high-calcium medium.
Open symbols correspond to low-calcium colonies, closed symbols to high-calcium colonies.
Symbol colors indicate colony cell number, given in the legend.
(\textit{I} and \textit{J}) Keratinocytes in low-calcium medium (\textit{I}) or after $24 \unit{h}$ in high-calcium medium (\textit{J}) labeled with anti-\hbox{E-cadherin} and anti-paxillin antibodies and stained with phalloidin to mark \hbox{F-actin}.
Scale bars in \textit{A}--\textit{D}, \textit{I}, and \textit{J} represent $50 \unit{\mu m}$.
Data for high-calcium colonies in \textit{F}--\textit{H} are adapted from~\cite{Mertz-2012}.
}
\end{figure*}

In general, low-calcium colonies exhibited traction stresses throughout the colony, usually pointing radially inward from the colony edge and in various directions in the interior (Fig.\ \ref{fig:Fig2}\textit{A}).
Regions of high strain energy were found throughout the interior (Fig.\ \ref{fig:Fig2}\textit{B}).
In contrast, high-calcium colonies displayed traction stresses generically pointing radially inward from the colony edge (Fig.\ \ref{fig:Fig2}\textit{C}) with hardly any strain energy beyond the colony edge (Fig.\ \ref{fig:Fig2}\textit{D}).

To quantify these spatial distributions, we plotted average strain energy density as a function of distance, $\Delta$, from the colony edge (as depicted in Fig.\ \ref{fig:Fig1}\textit{J}).
Average strain energy densities, $\bar{w}(\Delta)$, were normalized by the average strain energy density at the colony periphery, $\bar{w}(0)$.
These profiles (Fig.\ \ref{fig:Fig2}\ \textit{E} and \textit{F}) terminate where inward erosion covered the entire area of the colony, at $\Delta \approx R$, where $R$ is the effective radius of the colony, given by the radius of the disk with the same area as the colony.

In most low-calcium colonies, we observed some localization of strain energy at the colony periphery ($\Delta = 0$) and high amounts of strain energy throughout the colony ($\Delta > 0$), sometimes at the colony center ($\Delta \approx R$) (Fig.\ \ref{fig:Fig2}\textit{E}).
In contrast, the strain energy of nearly all the high-calcium colonies was strongly localized at the colony periphery, generally decaying to zero toward the colony center (Fig.\ \ref{fig:Fig2}\textit{F}).
Although this trend seems to hold regardless of  number of cells in the colony, the difference is much less pronounced for the smallest colonies ($R \lesssim 50\unit{\mu m}$).
The radii of small colonies are comparable to the traction-stress penetration length, $\ell_p$, which measures how far from the periphery traction stresses penetrate the colony.
Thus in small colonies, the stress measurements do not readily distinguish forces generated at the colony center or periphery.
In our previous study on high-calcium keratinocytes, we measured $\ell_p \approx 11 \unit{\mu m}$~\cite{Mertz-2012}.

Next, we quantitatively compared the spatial distributions of strain energy across these two colony populations with and without cadherin-based intercellular adhesions.
We calculated the total strain energy, $W$, exerted by each colony and the relative distance into the colony from its periphery, $\Delta/R$, required to capture 75\% of the total strain energy, $3W/4$.
We separated larger colonies ($R \gg \ell_p$, or $R>50\unit{\mu m}$) of the low- and high-calcium populations.
Large, low-calcium colonies required on average 10\% more inward erosion (statistically significant, $p=0.0002$) to achieve 75\% of the total colony strain energy than large, high-calcium colonies, whereas there was no significant difference in strain energy distribution for the populations of small ($R<50 \unit{\mu m}$) colonies (Fig.\ \ref{fig:Fig2}\textit{G}) ($p=0.43$).
These data suggest that formation of cadherin-based adhesions in high-calcium medium results in a shift in localization of traction stress from internal regions of the colony to the periphery.

The low- and high-calcium colonies did not seem to exhibit different amounts of average strain energy density.
A plot of total strain energy versus colony area, $A$, while scattered, shows no apparent difference between these populations~(Fig.\ \ref{fig:Fig2}\textit{H}).
In general, larger colonies tended to perform more work on the substrate.

\begin{figure*}[hbt]
\includegraphics[width=1.0\textwidth]{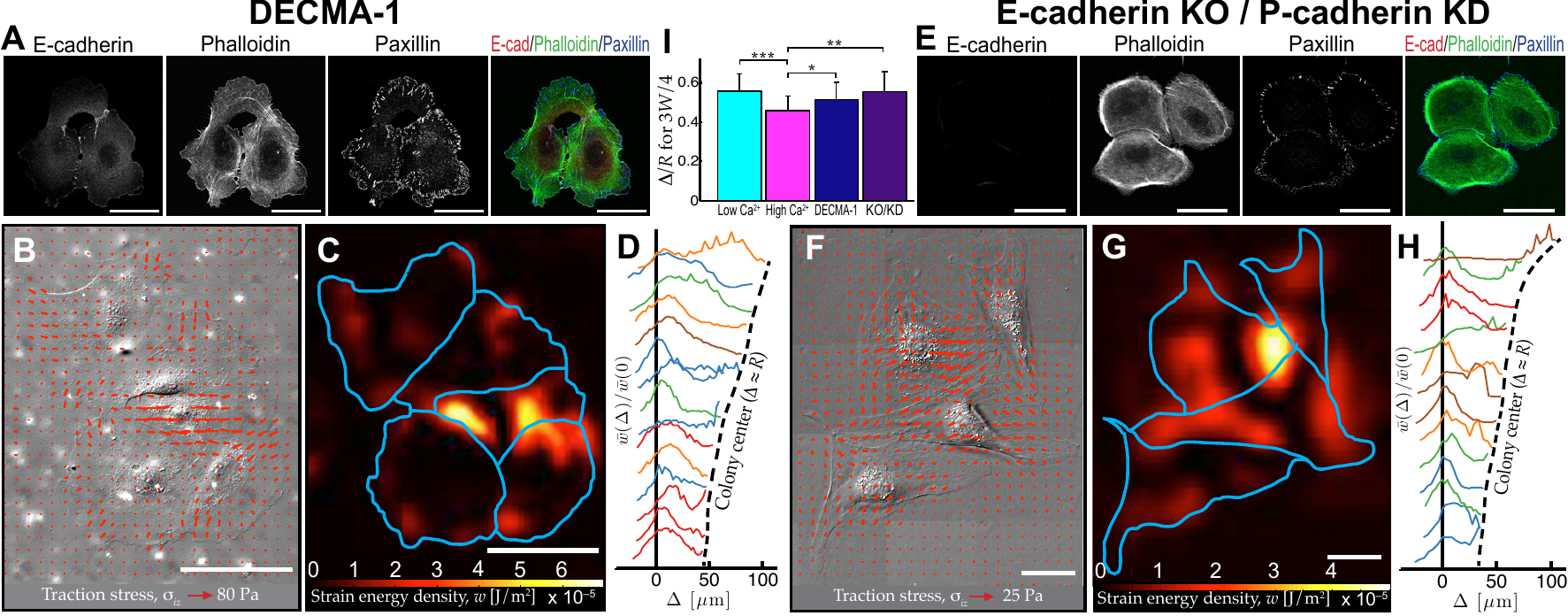}
\caption{\label{fig:Fig3} \textit{Cadherin-based adhesions are required for organization of traction stresses in high-calcium medium}.
(\textit{A}) Localization of \hbox{E-cadherin}, phalloidin (\hbox{F-actin}), and paxillin in colony of three wildtype keratinocytes in high-calcium medium for $24 \unit{h}$ with \hbox{DECMA-1}.
(\textit{B}) Distribution of traction stresses (red arrows) of five-cell colony in high-calcium medium for $24 \unit{h}$ with \hbox{DECMA-1} overlaid on DIC image of colony.
For clarity, one-sixteenth of calculated traction stresses are shown.
(\textit{C}) Strain energy of colony in \textit{B} with individual cell outlines in blue.
(\textit{D}) Strain energy profiles for $n=15$ \hbox{DECMA-1}-treated colonies.
Each solid curve represents colony's average strain energy density as a function of distance, $\Delta$, from colony the edge, as defined in Fig.\ \ref{fig:Fig1}\textit{J}.
For clarity, profiles are spaced vertically according to colony size.
Each profile terminates where inward erosion covers entire colony area, at $\Delta \approx R$.
(\textit{E}) Localization of \hbox{E-cadherin}, phalloidin (\hbox{F-actin}), and paxillin in a colony of three \hbox{E-cadherin}-knockout/\hbox{P-cadherin}-knockdown (KO/KD) keratinocytes after $24 \unit{h}$ in high-calcium medium.
(\textit{F}) Distribution of traction stresses (red arrows) of a colony of three KO/KD keratinocytes in high-calcium medium for $24 \unit{h}$ overlaid on DIC image of colony.
For clarity, one-sixteenth of calculated traction stresses are shown.
(\textit{G}) Strain energy distribution of colony in \textit{F} with individual cell outlines in blue.
(\textit{H}) Strain energy profiles for $n=14$ KO/KD colonies after $24 \unit{h}$ in high-calcium medium.
As in \textit{D}, profiles were calculated as defined in Fig.\ \ref{fig:Fig1}\textit{J}.
Profile colors in \textit{D} and \textit{H} correspond colony cell number given by the legend between Fig.~\ref{fig:Fig2}\ \textit{E} and \textit{F}.
(\textit{I}) Comparison of the strain energy distribution for large low-calcium ($n=24$), large high-calcium ($n=21$), \hbox{DECMA-1} ($n=15$), and KO/KD ($n=14$) colonies.
Values represent proportion of the colony from periphery inward, $\Delta /R$, necessary to capture 75\% of total colony strain energy.
Error bars indicate one standard deviation.
Higher proportions indicate higher strain energy nearer colony center.
Statistical significance between pairs of colony conditions is indicated by one asterisk ($\ast$) ($p<0.05$), two asterisks ($\ast$$\ast$) ($p<0.01$), or three asterisks ($\ast$$\ast$$\ast$) ($p<0.001$).
Scale bars in \textit{A}--\textit{C} and \textit{E}--\textit{G} represent $50 \unit{\mu m}$.
}
\end{figure*}

Because low- and high-calcium keratinocyte colonies have different arrangements of cytoskeletal and adhesion proteins, we characterized spatial localizations of actin, \hbox{E-cadherin}-mediated cell--cell adhesions, and focal adhesions in keratinocyte colonies using phalloidin staining and immunohistochemistry (\textit{Materials \& Methods}).
\hbox{E-cadherin} is highly expressed in keratinocytes, mediates adhesive activity, and is essential for adherens-junction formation.
In high-calcium colonies, \hbox{E-cadherin} was localized at keratinocyte junctions~(Fig.\ \ref{fig:Fig2}\textit{I}).
Positions of actin stress fibers were correlated with areas of strong \hbox{E-cadherin} localization, and there was coordination of the orientation of actin fibers across multiple cells, consistent with earlier reports on cytoskeletal rearrangement after calcium elevation~\cite{Vaezi-2002}.
While traction stresses of low- and high-calcium colonies had different spatial distributions, focal adhesions, marked by paxillin, were concentrated at the colony periphery in both cases.

\subsection{Cadherin-Based Adhesions are Required for Organization of Traction Stresses in High-Calcium Medium}
Because elevation of extracellular calcium modulates cellular properties in addition to cadherin-based-adhesion induction~\cite{Pillai-1990,Micallef-2009}, we sought to isolate the role of cadherin in spatially organizing traction forces.
We used two different methods to inhibit formation of cadherin-based adhesions in the presence of high-calcium medium.
First, we used the function-blocking antibody \hbox{DECMA-1}, which prevents homophilic binding between extracellular domains of \hbox{E-cadherin}~\cite{Hordijk-1997}.
\hbox{DECMA-1} was added to keratinocyte colonies with high-calcium medium for $24 \unit{h}$.
Immunostaining of these colonies showed strong reduction of \hbox{E-cadherin} at intercellular contact (Fig.\ \ref{fig:Fig3}\textit{A}).
Despite this change, we observed minimal coordination of actin across multiple cells in a colony, and focal adhesions were present at the colony periphery and throughout the colony interior.
In keratinocytes in high-calcium medium with \hbox{DECMA-1}, we measured traction stress and strain energy throughout the colony, in particular at cell--cell contacts (Fig.\ \ref{fig:Fig3} \textit{B} and \textit{C}).
Strain energy profiles of 15 \hbox{DECMA-1}-treated colonies (all with $R>50 \unit{\mu m}$) show many cases of high strain energy transmitted in the colony interior (Fig.\ \ref{fig:Fig3}\textit{D}).

We further investigated the role of classical cadherins using primary keratinocytes from an epidermal-\hbox{E-cadherin}-knockout~(KO) mouse~\cite{Tunggal-2005}.
We used shRNA to knock down~(KD) the other classical cadherin expressed in these cells, \hbox{P-cadherin}, which is upregulated in \hbox{E-cadherin}-null cells~\cite{Michels-2009} (\textit{Materials \& Methods}).
We analyzed cell--cell and cell-matrix adhesions by immunostaining KO/KD cells cultured in high-calcium medium for $24 \unit{h}$.
Colonies of KO/KD cells showed no \hbox{E-cadherin} expression, did not coordinate their actin cytoskeletons across multiple cells, and displayed a slight reduction of focal adhesions underneath the colony (Fig.\ \ref{fig:Fig3}\textit{E}).
As with \hbox{DECMA-1}-treated colonies, KO/KD colonies in high-calcium medium for $24 \unit{hr}$ showed traction stresses and strain energy underneath cell--cell contacts (Fig.\ \ref{fig:Fig3}\ \textit{F} and \textit{G}).
Strain energy profiles of 14 KO/KD colonies in high-calcium medium (all with $R>50 \unit{\mu m}$) show strong strain energy transmitted throughout the colony (Fig.\ \ref{fig:Fig3}\textit{H}). 

\begin{figure}[hbt]
\centering
\includegraphics[width=0.5\textwidth]{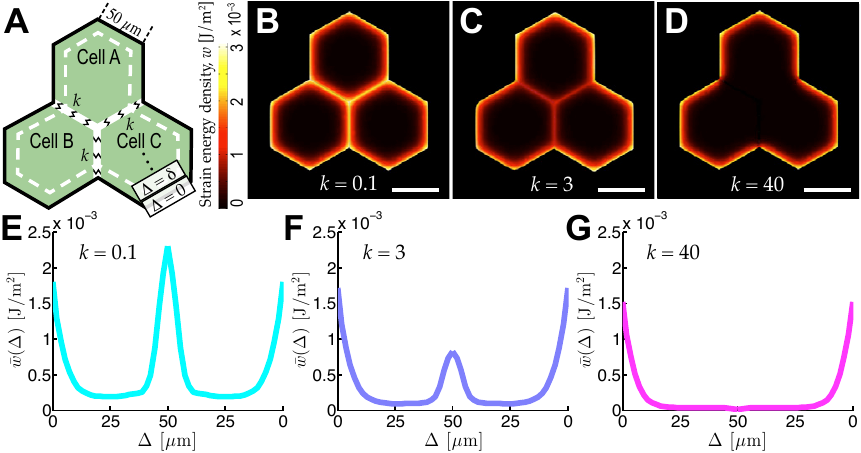}
\caption{\label{fig:Fig4} \textit{Minimal physical model captures cadherin-dependent organization of traction stresses}.
(\textit{A}) Schematic of planar colony of three hexagonal cells.
(\textit{B}--\textit{D}) Strain energy distributions for colony of three hexagonal cells with different spring stiffness, $k$, expressed in units of $E/L$, where $E$ is the Young's modulus of the cell and $L$ the side length of each hexagon.
(\textit{E}--\textit{G}) Spatial profiles of average strain energy as a function of distance, $\Delta$, from colony edge for different values of $k$ corresponding to data in \textit{B}--\textit{D}.
Other parameters: $\ell_p/L=0.2$, $E=1 \unit{kPa}$, $\nu=0.4$, $\sigma_a=4 \unit{kPa}$, $h=0.2 \unit{\mu m}$, $Y=2\times10^6 \unit{N/m^3}$ (\textit{Appendix}).
Scale bars in \textit{B}--\textit{D} represent $50 \unit{\mu m}$.
}
\end{figure}

DECMA-1-treated colonies needed on average 6\% more inward erosion than large high-calcium wildtype colonies to achieve 75\% of the total colony strain energy (statistically significant, $p=0.048$).
KO/KD colonies required on average 10\% more inward erosion than large high-calcium wildtype colonies to achieve 75\% of the total colony strain energy (statistically significant, $p=0.002$) (Fig.\ \ref{fig:Fig3}\textit{I}).
Compared to large low-calcium colonies using this same measure, neither \hbox{DECMA-1}-treated colonies ($p=0.14$) nor KO/KD colonies ($p=0.94$) showed significant differences in spatial distributions of strain energy.
Thus, keratinocytes in high-calcium medium are unable to organize traction forces to the colony periphery in the absence of cadherin-based cell--cell junctions.

\subsection{Minimal Physical Model Captures Cadherin-Dependent Organization of Traction Stresses}
Because of the simple spatial trends of traction stresses observed in colonies with and without intercellular adhesions, we examined whether a minimal physical model could reproduce the experimental results.
We model each cell in a colony as a homogeneous and isotropic elastic material~\cite{Edwards-2011,Banerjee-2011}.
In our model, each cell exerts a contractile ``pressure'' opposed by strong adhesion to a compliant substrate~\cite{Banerjee-2012}.
At each point within a cell, we require that these opposing forces balance.
This model ignores all active processes modulated by cell--cell adhesions, including downstream signaling, and represents each intercellular adhesion as a purely physical connection characterized by a spring constant, $k$~\cite{Notbohm-2012}.

To make predictions with this model, we use a numerical solution of the two-dimensional governing equations (\textit{Appendix}).
To mimic the cell geometry in the timecourse experiment (Fig.\ \ref{fig:Fig1}), we consider the case of three hexagonal cells (Fig.\ \ref{fig:Fig4}\textit{A}).
We find that, for increasing cell--cell-coupling strength, $k$, traction stress and strain energy disappear under cell--cell junctions (Fig.\ \ref{fig:Fig4}\ \textit{B--D}), recapitulating the transition seen in real cells stimulated by calcium elevation (Fig.\ \ref{fig:Fig1}\ \textit{D--F}).
The similarity between model and experiment is also evident in plots of strain energy density as a function of distance from the colony edge (Fig.\ \ref{fig:Fig4}\ \textit{E--G} and Fig.\ \ref{fig:Fig1}\textit{K}).

The model demonstrates the importance of intercellular-adhesion strength in spatially organizing cell-ECM forces.
For weak cell--cell coupling (small $k$), individual cells deform the substrate independently of each other, with significant substrate deformation at all edges of each cell.
On the other hand, strongly coupled colonies (large $k$) behave as a cohesive, contractile unit, with substrate deformation only at the colony periphery.

This planar model is an extension of an analytically tractable, one-dimensional model (\textit{Appendix}).

\section{Discussion}
Our results show that cadherin-based cell--cell adhesions modulate force transmission to the ECM.
In particular, our traction-force data on cohesive cell colonies suggest that intercellular-adhesion formation through classical cadherins reorganize the spatial distributions of traction stress.
In colonies of cells with strong \hbox{E-cadherin}-based adhesions, cell-ECM traction stresses are localized in a ring around the colony periphery.
In weakly cohesive colonies, regions of high traction stress appear throughout the colony.
Furthermore, traction stresses cannot reorganize in high-calcium medium when cadherin-based adhesion is inhibited.
Comparison of our experimental data with our minimal physical model suggests that strong physical cohesion between cells is sufficient to drive the relocalization of cell-ECM forces to the periphery of cell colonies.
While our data show that E-cadherin is necessary to reorganize traction forces, \hbox{E-cadherin} alone may not be sufficient.
Further study is required to determine whether additional adhesive processes downstream of adherens junctions, such as the formation of desmosomes by nonclassical cadherins~\cite{Michels-2009}, are necessary to achieve sufficient cohesion.

Our findings resonate with recent studies on cellular adhesion pointing toward crosstalk of cadherin- and integrin-based adhesions.
Focal adhesions have been observed to disappear underneath cell--cell contacts~\cite{Yamada-2007,leDuc-2010}, but this effect may depend on substrate stiffness~\cite{McCain-2012} and the extent of cell spreading~\cite{Nelson-2004}.
Recent work has also suggested that forces transmitted through focal adhesions can modulate intercellular forces~\cite{Maruthamuthu-2011,McCain-2012}, which in turn can modulate intercellular-junction assembly and disassembly~\cite{Liu-2010}.
Our study highlights intercellular adhesions' ability to impact cell-ECM force generation, which allows for bidirectional feedback between cell--cell and cell-matrix forces.
Indeed, tension at cadherin junctions~\cite{Vasioukhin-2000,Vaezi-2002} is known to elicit cell-signaling events and actin dynamics~\cite{Potard-1997,Bard-2008,leDuc-2010,Yonemura-2010,Niessen-2011} and contribute to collective cell migration~\cite{Tambe-2011,Weber-2012}.
In light of these prior results on integrin-cadherin feedback, it is somewhat surprising that a minimal physical model can capture the observed dependence of cell-matrix forces on the strength of cadherin-mediated cell--cell adhesions.

Reorganization of cell-ECM forces is likely one important mechanism by which cadherin-based adhesions drive tissue morphogenesis and homeostasis.
In development, differential adhesion has been shown to play an important role in cell sorting~\cite{Steinberg-1963,Kane-2005,Steinberg-2007}, and the reorganization of intercellular forces in this context is entirely unexplored.
Furthermore, in wound healing, we expect strong cell-ECM forces to be generated at a wound edge due to the local loss of intercellular adhesion.
These forces could act as a signal, inducing migratory behavior in epithelial cells~\cite{Trepat-2009,Khalil-2010}, activating responses of stromal cells, and organizing the ECM~\cite{Raghavan-2000,Grinnell-2000,Dzamba-2009}.
A key avenue for future investigations will be to explore how organization of force stimulates cellular responses within tissues.

\vspace{8 mm}
\noindent\scriptsize{\textbf{ACKNOWLEDGMENTS.}
We are grateful to Margaret L.\ Gardel (University of Chicago) and Alpha S.\ Yap (University of Queensland) for helpful discussions.
We thank Barbara Boggetti (University of Cologne) for preparation of KO/KD cells.

This work was supported by a National Science Foundation Graduate Research Fellowship to A.F.M., German Cancer Aid and Sonderforschungsbereich (829~A1 and Z2) grants to C.M.N., and National Science Foundation grants to M.C.M. (DMR-0806511 and DMR-1004789) and to E.R.D. (DBI-0619674).
V.H. is a Pew Scholar in Biomedical Research and is funded by the National Institutes of Health (AR060295) and the Connecticut Department of Public Health (\hbox{12-SCB-YALE-01}).}

\section{Materials \& Methods}
\scriptsize{
\textbf{Preparation of Substrates for Traction Force Microscopy.}
A borate buffer solution was made from deionized water with $3.8 \unit{mg/ml}$ sodium tetraborate and $5 \unit{mg/ml}$ boric acid. 
Silane (3-aminopropyl triethoxysilane) (Polysciences) was vapor-deposited onto $35 \unit{mm}$ glass-bottom dishes (WillCo Wells) to allow florescent beads to be bonded to the surface.
Beads were deposited by filling the dish with a solution containing dark-red florescent (660/680) carboxylate-modified microspheres with radius $0.1\unit{\mu m}$ (Life Technologies) at a volume ratio of 1:3,000 and 1 wt\% 1-ethyl-3-(3-dimethylaminopropyl)carbodiimide (EDC) (Sigma-Aldrich) at a volume ratio of 1:100 in borate buffer.
Silicone elastomer was then prepared by mixing a 1:1 weight ratio of CY52-276A and CY52-276B (Dow Corning Toray).
After being degased for $10\unit{min}$, the elastomer was spin-coated onto the glass of the dish at $2$,$000\unit{rpm}$ for $60\unit{s}$.
The dish was cured at $50^{\circ}$C for $3 \unit{min}$ and resulted in an elastic film $\sim$$21\unit{\mu m}$ thick.
With the elastomer cross-linked, silane was vapor-deposited on the elastomer-coated dish.
A second layer of florescent polystyrene beads was deposited at a higher concentration, volume ratios of 1:1,000 beads and 1:100 EDC in borate buffer.
A second layer of fresh, degased elastomer was spin-coated at $10$,$000 \unit{rpm}$ for $90 \unit{s}$ resulting in a layer $\sim$$3 \unit{\mu m}$ thick.
The sample was cured at RT overnight.
We estimated the Young's modulus, $E$, of the cured elastomer to be $\sim$$3 \unit{kPa}$ using bulk rheology.
Before cells were plated, the elastomer surface was coated with fibronectin from bovine plasma (Sigma-Aldrich) at a concentration of $0.2\unit{mg/ml}$, which sat for $20 \unit{min}$ at RT before being washed off with PBS.

\textbf{Confocal Microscopy.}
Images for TFM experiments were acquired using an Andor Revolution spinning-disk confocal system (Andor Technology) mounted on an inverted microscope (Nikon Eclipse Ti) with a Plan Apo 60$\times$ water-immersion objective lens with numerical aperture of 1.2 (Nikon).
A $640 \unit{nm}$ laser and DIC channel were used to image florescent beads and cells, respectively.
Images were acquired with an iXon EMCCD camera with a resolution of $1$,$024 \times 1$,$024$ pixels (Andor Technology).
The field of view was $113 \times 113 \unit{\mu m^2}$.
Because a single field of view was too small to image an entire cell colony, between 9 and 42 fields of view per colony were acquired, with adjacent fields of view overlapping by $\sim$$25\%$ and stitched together with sub-pixel precision by aligning bead positions in overlapping regions.
The stage was controlled through Motion Controller/Driver SMC100CC high-speed motorized actuators (Newport).
We imaged fluorescent beads with confocal image stacks of total thickness $5 \unit{\mu m}$ to cover the beads' entire point-spread function in $z$.
Confocal image slices were spaced $200 \unit{nm}$ apart.
These stacks were reduced to single images for particle-tracking by averaging the slices from five below to five above the slice with the highest total intensity.

\textbf{Live-Cell Imaging.}
Confocal image stacks of the fluorescent beads were acquired for each cell condition.
Control images of the beads in their unstressed state were acquired after removing the cells from the elastomer with proteinase~K (Life Technologies) at $0.5 \unit{mg/ml}$ for $5 \unit{min}$ and then washing with PBS.
The cells on the microscope were maintained at $37^{\circ}$C using a heated microscope stage.
pH was controlled using HEPES solution at $15 \unit{mM}$ (Sigma-Aldrich).
To inhibit the formation of cadherin-based adhesions, we added anti-\hbox{E-cadherin} antibody \hbox{DECMA-1} (\mbox{Abcam}) at $6 \unit{\mu g/ml}$ to the high-calcium medium.
For consistency across cellular conditions, we controlled for colonies that deviated significantly from disk-shaped and contained cells with long protrusions by selecting for colonies whose actual perimeter, $P$, was no more than 1.5 times the perimeter of a circle of the same area, $A$, as the colony ($P \leq 3\sqrt{\pi A}$).

\textbf{Calculation of Traction Stresses and Strain Energies.}
After determining bead positions using centroid analysis in \textsc{matlab}~\cite{Crocker-1996}, we calculated the deformation of the substrate, $u^s_i({\bf r},z_\text{o})$, across its stressed (with cells) and unstressed (with cells removed) states, where $z_\text{o}$ is the distance between the substrate bottom and the bead layer.
In Fourier space, the deformation field is related to the traction stresses at the surface of the substrate, $h_s$, via linear elasticity, $\sigma^s_{iz}({\bf k},h_s) = Q_{ij}({\bf k},z_\text{o},h_s) u^s_j({\bf k},z_\text{o})$, where ${\bf k}$ represents the in-plane wave vector.
Here, $\sigma^s_{iz}({\bf k},h_s)$  and $u^s_j({\bf k},z_\text{o})$ are the Fourier transforms of the traction stress on the top surface and the displacements of the bead layer just below the surface, respectively.
The tensor, $Q$, depends on the thickness and modulus of the substrate, the location of the beads, and the wave vector \cite{delAlamo-2007, Xu-2010}.
We calculated the strain energy density, $w({\bf r})=\frac{1}{2}\sigma^s_{iz}({\bf r},h_s) u^s_i({\bf r},h_s)$~\cite{Butler-2001}.
The deformation on the surface was determined using $u^s_{i}({\bf k},h_s)=Q_{ij}^{-1}({\bf k},h_s,h_s) Q_{jk}({\bf k},z_{\text{o}},h_s)u^s_k({\bf k},z_{\text{o}})$.
Because of their small size and immersion in a viscous medium, we expect the colonies of cells to be in mechanical equilibrium (net force of zero).  
Due to experimental error in determining substrate displacement fields, we occasionally calculated non-zero net traction forces on a colony.
We discarded colonies with more than 15\% residual force,
$\left| \int dA \left(\sigma_{xz}^{s} \hat{\bm{\textrm{x}}} +\sigma_{yz}^s \hat{\bm{\textrm{y}}} \right) \right| \geq 0.15 \int dA \left|\sigma_{xz}^{s} \hat{\bm{\textrm{x}}} +\sigma_{yz}^s \hat{\bm{\textrm{y}}} \right|$.

\textbf{Primary Keratinocyte Culture.}
Primary wildtype keratinocytes were isolated as described~\cite{Nowak-2009}.
Briefly, isolated backskin of newborn CD1 mice was floated on dispase overnight at $4^{\circ}$C.
The epidermis was separated from the dermis with forceps and incubated in 0.25\% trypsin for $10 \unit{min}$ at RT.
Individual cells were released by trituration and plated on \hbox{mitomycin-C}-treated J2 fibroblasts in low-calcium medium ($0.05 \unit{mM}$ CaCl$_2$).
After 2--4 passages, cells were plated on plastic dishes without feeder cells.
Primary keratinocytes were also isolated as described~\cite{Pasparakis-2002} from newborn epidermis in which \hbox{E-cadherin} was conditionally deleted as described~\cite{Tunggal-2005}.
KO/KD cells were generated by lentiviral transduction of \hbox{E-cadherin}-deficient keratinocytes using shRNA directed against P-cadherin, as described~\cite{Michels-2009}.
Cadherin-junction formation was induced by raising the concentration of CaCl$_2$ of the low-calcium medium to $1.5 \unit{mM}$.

\textbf{Immunohistochemistry.}
Cells were fixed in 3.7\% formaldehyde for $10 \unit{min}$ and then washed twice for $2 \unit{min}$ in PBS.
A blocking solution of normal goat serum, normal donkey serum, bovine serum albumin, gelatin, and triton~X in PBS was used to prevent non-specific binding.
Cells were stained using $3$,$000 \unit{units/\mu l}$ Alexa Fluor 594 phalloidin (Life Technologies) and primary antibodies $8 \unit{ng/\mu l}$ monoclonal mouse anti-\hbox{E-cadherin} (TaKaRa) and $4 \unit{ng/\mu l}$ rabbit anti-paxillin (Sigma-Aldrich).
After being washed in PBS, cells were incubated with secondary antibodies $8 \unit{ng/\mu l}$ goat anti-rabbit Alexa Fluor 488 (Life Technologies) and $8 \unit{ng/\mu l}$ goat anti-rat Alexa Fluor 647 (Life Technologies) and again with $3$,$000 \unit{units/\mu l}$ Alexa Fluor 594 phalloidin.
Cells were then mounted in ProLong Gold with DAPI (Life Technologies).

Fluorescent images of immunohistochemical staining were acquired using confocal laser scanning microscopy on a Zeiss LSM 510 system equipped with Ar, HeNe 543, and HeNe 633 laser lines allowing imaging with lasers of wavelengths $488$, $568$, and $633 \unit{nm}$ and a Plan Apo 40$\times$ oil-immersion objective with numerical aperture 1.3 (Zeiss).
The field of view was $313 \times 313 \unit{\mu m^2}$ with a maximum resolution of $2$,$048 \times 2$,$048$ pixels.
The stage was controlled using an MCU 28 unit (Zeiss).

\textbf{Statistical Analyses.}
Statistical significance for strain-energy distributions was assessed with $p$ values determined by two-sided Student's $t$ tests.
Statistically significant $p$ values were those lower than 0.05. 
}

\begin{figure}[hbt]
\includegraphics[width=0.5\textwidth]{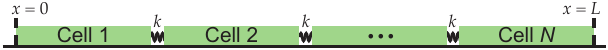}
\caption{Minimal one-dimensional picture of $N$ cells adhering via cadherin-based adhesions, modeled as linear springs of stiffness $k$.
}
\label{fig:image}
\end{figure}

\section{Appendix}
\subsection{Planar Model of Cell Colonies as Elastic Media}

\small{
We model each cell in a colony as a contractile elastic object, assumed to be homogeneous and isotropic with Young's modulus, $E$, and Poisson's ratio, $\nu$~\cite{Edwards-2011,Banerjee-2011}.
The constitutive relation for the thickness-averaged cellular stress tensor, $\sigma_{ij}$, with $i$, $j$ denoting in-plane coordinates, is given by
\begin{equation}\label{eq:stress2D}
\sigma_{ij}=\frac{E}{2(1+\nu)}\left(\frac{\nu}{1-\nu} {\bm \nabla}\cdot{\bf u}\ \delta_{ij} + \partial_i u_j + \partial_j u_i\right) + \sigma_a \delta_{ij},
\end{equation}
where ${\bf u}$ is the cellular displacement field, $\delta_{ij}$ is the Kronecker delta, and $\sigma_a>0$ denotes the active stress due to actomyosin contractility.
In the thin-film limit, force-balance requires
\begin{equation}
\label{eq:forcebalance}
h\partial_j \sigma_{ij} = Y u_i,
\end{equation}
where $h$ is the average cell height, and $Y$ is a parameter describing the strength of cell-substrate coupling and depends on substrate stiffness as well as on the strength of focal adhesions~\cite{Edwards-2011,Banerjee-2011,German-2012,Banerjee-2012}. 

We model cell--cell interactions as linear springs~\cite{Notbohm-2012} of spring constant $k$ per unit area, exerting a harmonic force ${\bf f}$ per unit area normal to the interface between two cells.
The addition of springs translates into boundary conditions at the intercellular interfaces as $\sigma_{ij} n_j=f_i$, with ${\bf n}$ denoting the outward unit normal.
The edge of the colony, however, respects the stress-free boundary condition, $\sigma_{ij}n_j=0$.
We numerically solve Eqs.~\eqref{eq:stress2D} and \eqref{eq:forcebalance} subject to the aforementioned boundary conditions using the \textsc{matlab} PDE Toolbox.
We evaluate strain energy density, $w$, given by $w=\frac{1}{2} {\bf T}\cdot {\bf u}$, where ${\bf T} = Y{\bf u}$ is the local traction stress exerted by the colony.
According to this model, spatial variation of traction stresses and strain energy densities in the direction normal to the edges is characterized by a penetration length, $\ell_p=\sqrt{E h\nu/[Y(1-\nu^2)]}$.

\subsection{Model of Cell Colonies as Elastic Media in One Dimension}
As in the planar case, individual cells are described in one dimension as thin active elastic materials adherent to an elastic substrate.
We consider $N$ cells, each of rest length $L/N$ and average height $h$, with cell--cell adhesions modeled by linear springs of stiffness $k$ (Fig.~\ref{fig:image}).
Let $\sigma^{(\alpha)}$ denote the internal stress in the $\alpha^{\textrm{th}}$ cell and $u^{(\alpha)}$ the corresponding displacement field.
The one-dimensional equivalents of Eqs.~\eqref{eq:stress2D} and \eqref{eq:forcebalance} for the $\alpha^{\textrm{th}}$ cell are given by
\begin{equation}
\sigma^{(\alpha)}(x)=B\partial_x u^{(\alpha)} + \sigma_a
\end{equation}
and
\begin{equation}
h \partial_x \sigma^{(\alpha)}=Yu^{(\alpha)},
\end{equation}
respectively, where $B$ is the longitudinal elastic modulus of the cell.
In this one-dimensional picture, substrate-induced traction penetration length is given by $\ell_p=\sqrt{B h/Y}$.

\begin{figure*}[hbt]
\includegraphics[width=0.99\textwidth]{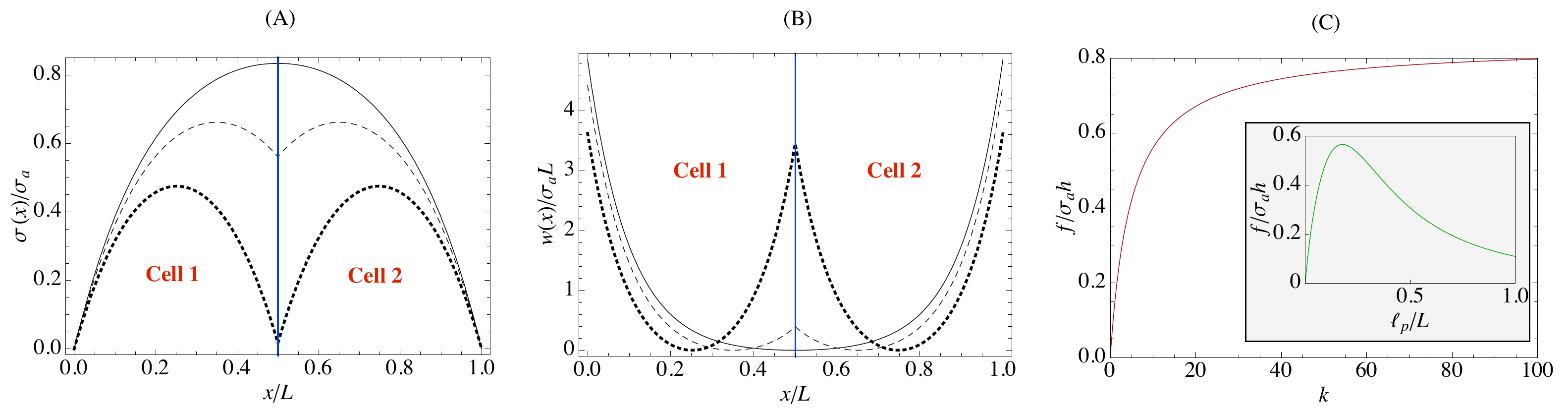}
\caption{(\textit{A}) Internal stress and (\textit{B}) strain energy density in a one-dimensional adherent cell-pair for $kL/\sigma_a=0.004$ (dotted), $kL/\sigma_a=0.4$ (dashed), and $kL/\sigma_a=40$ (solid). Parameters: $\ell_p/L=0.2$, $B/\sigma_a=2$, $h/L=0.1$.
\textit{(C)} Intercellular force, $f$ versus intercellular adhesion strength, $k$ (in units of $\sigma_a/L$) for $\ell_p/L=0.2$.
Inset: $f$ as a function of $\ell_p/L$ for $kL/\sigma_a=10$.
}
\label{fig:1dmodel}
\end{figure*}

Internal stress distribution in the colony is then governed by equations
\begin{equation}
\ell_p^2\partial_x^2\sigma^{(\alpha)}(x) + \sigma_a = \sigma^{(\alpha)}(x)\ \textrm{for}\ \ 1\leq \alpha \leq N
\end{equation}
subject to boundary conditions
\begin{subequations}
\begin{align}
\sigma^{(1)}\vert_{x=0} &= 0,\\
\sigma^{(\alpha)}\vert_{x=\alpha L/N} &=  \sigma^{(\alpha+1)}\vert_{x=\alpha L/N} \notag \\
&= k\left[ u^{(\alpha+1)}-u^{(\alpha)}\right]_{x=\alpha L/N}\ \textrm{for}\ \ 1\leq \alpha<N,\\
\sigma^{(N)}\vert_{x=L} &= 0.
\end{align}
\end{subequations}
For simplicity, we assume that the cell--cell adhesion springs have zero rest length and that the colony ends ($x=0,L$) respect stress-free boundary conditions.

Explicit solutions for cellular stresses in an adherent cell-pair ($N=2$) are given by
\begin{subequations}
\begin{align}
\sigma^{(1)}(x) /\sigma_a =&\; 1 - \exp{\left(-\frac{x}{\ell_p}\right)}  \\
&+  \frac{2\sinh{\left(\frac{x}{\ell_p}\right)}\left[\frac{2k\ell_p}{B}+\exp{\left(\frac{L}{2\ell_p}\right)}-1\right]}{\frac{2k\ell_p}{B}\left[1+\exp{\left(\frac{L}{\ell_p}\right)}\right]+\exp{\left(\frac{L}{\ell_p}\right)}-1}, \notag \\
\sigma^{(2)}(x)/\sigma_a =&\; 2 \sinh{\left(\frac{L-x}{2\ell_p}\right)}\\
& \times \frac{\cosh{\left(\frac{x}{2\ell_p}\right)}-\cosh{\left(\frac{L-x}{2\ell_p}\right)} +\frac{2k\ell_p}{B}\sinh{\left(\frac{x}{2\ell_p}\right)}}{\frac{2k\ell_p}{B}\cosh{\left(\frac{L}{2\ell_p}\right)}+\sinh{\left(\frac{L}{2\ell_p}\right)}}.  \notag
\end{align}
\end{subequations}
For weak intercellular coupling, $k\ll B/\ell_p$, internal stresses are maximal at the center of individual cells and negligible at the cell--cell junction.
For a strongly coupled cell-pair, $k\gg B/\ell_p$, internal stresses build up at the junction between the cells, which corresponds to the limit of a cohesive cell colony (Fig.~\ref{fig:1dmodel}\textit{A}).
In this case, internal stress takes the simple form
\begin{equation}
\frac{\sigma(x)}{\sigma_a} = 1 - \frac{\cosh\left(\frac{L-2x}{2\ell_p}\right)}{\cosh\left(\frac{L}{2\ell_p}\right)}.
\end{equation}
Strain energy density, $w$, in the cell-pair is determined using $w(x)=\frac{1}{2}T(x)u(x)$, where the traction $T(x)=Yu(x)$.
For a weakly coupled cell-pair ($k\rightarrow 0$), $w$ is localized at the edges of individual cells, and the net traction force on each individual cell vanishes.
In contrast, a strongly cohesive cell-pair ($k \rightarrow \infty$) behaves as a single cell, with strain energy density localized at the edge of the pair and vanishing at the junction (Fig.~\ref{fig:1dmodel}\textit{B}).
For intermediate strengths of cell--cell adhesion, there is finite but small strain energy at the junction compared to the edges of the cell-pair.
Traction force imbalance at each cell gives the estimate of the total force, $f$, transmitted to the intercellular adhesion,
\begin{eqnarray}
f & = & \left| \int_0^{L/2} dx\ T^{(1)}(x) \right| \notag \\
& = & h\left| \sigma(0)-\sigma(L/2)\right| \notag \\
& = & \frac{h \sigma_a \left|1-\cosh{\left(\frac{L}{2\ell_p}\right)}\right|}{\cosh{\left(\frac{L}{2\ell_p}\right)} + \frac{B}{2k\ell_p} \sinh{\left(\frac{L}{2\ell_p}\right)}} \\
& \simeq & h\sigma_a\frac{k}{k + B/2\ell_p}\ \textrm{for}\ \ L\gg \ell_p. \notag
\end{eqnarray}
Intracellular force, $f$, grows monotonically with adhesion strength, $k$, before reaching a plateau when $k$ is large (Fig.~\ref{fig:1dmodel}\textit{C}).
However, the dependence of $f$ on penetration length, $\ell_p$, which is inversely related to substrate stiffness, is non-monotonic (Fig.~\ref{fig:1dmodel}\textit{C}, inset).
This biphasic relation arises from the competition among different elastic components (cell, substrate, and the intercellular spring) connected in series.
For small $\ell_p$, the substrate is deformed less compared to the cells and to the intercellular spring, leading to a rise in intercellular force.
A more compliant substrate with large $\ell_p$ is likely to accommodate larger cellular forces, reducing the net force transmitted to the intercellular adhesion.
}

\end{document}